\documentclass[12pt]{iopart}
\usepackage{amssymb}

\usepackage{iopams}
\usepackage{epsf}

\begin{document}

\title{Phase-transition-like Behavior of Quantum Games}

\author{Jiangfeng Du\footnote[1]{To whom correspondence should be addressed. E-mail: djf@ustc.edu.cn}}
\address{Department of Modern Physics, University of Science and Technology of China,
Hefei, 230027, People's Republic of China}
\address{Department of Physics, National University of Singapore, Lower Fent Ridge,
Singapore 119260, Singapore}
\address{Centre for Quantum Computation, Department of Applied Mathematics and
Theoretical Physics, University of Cambridge, Wilberforce Road, Cambridge CB3 0WA, United Kingdom}

\author{Hui Li\footnote[2]{E-mail: lhuy@mail.ustc.edu.cn}}
\address{Department of Modern Physics, University of Science and Technology of China,
Hefei, 230027, People's Republic of China}

\author{Xiaodong Xu}
\address{Harrison M. Randall Laboratory of Physics, The University of Michigan, Ann
Arbor, Michigan 48109-1120}

\author{Xianyi Zhou and Rongdian Han}
\address{Department of Modern Physics, University of Science and Technology of China,
Hefei, 230027, People's Republic of China}

\begin{abstract}
The discontinuous dependence of the properties of a quantum game on its
entanglement has been shown up to be very much like phase transitions viewed
in the entanglement-payoff diagram [J. Du \textit{et al.}, Phys. Rev. Lett,
\textbf{88}, 137902 (2002)]. In this paper we investigate such
phase-transition-like behavior of quantum games, by suggesting a method
which would help to illuminate the origin of such kind of behavior. For the
particular case of the generalized Prisoners' Dilemma, we find that, for
different settings of the numerical values in the payoff table, even though
the classical game behaves the same, the quantum game exhibits different and
interesting phase-transition-like behavior.
\end{abstract}

\pacs{03.67.-a, 02.50.Le}

\submitto{\JPA}

\maketitle

\section{Introduction}

The theory of quantum games is a new born field which combines the classical
game theory and the quantum information theory, opening a new range of
potential applications. Recent research have shown that quantum games can
outperform their classical counterparts \cite{1,2,3,4,5,6,11,7,8,15,10}. J.
Eisert \textit{et al}. investigated the quantization of the famous game of
Prisoners' Dilemma \cite{4}. Their result exhibits the surprising superiority
of quantum strategies over classical ones and the players can escape the
dilemma when they both resort to quantum strategies. L. Marinatto and T. Weber
studied the quantum version of the Battle of the Sexes game and found that the
game can have a unique solution with entangled strategy \cite{5}. Besides two
player quantum games, works on multiplayer games have also been presented
\cite{6,11}. In a recent paper of S.C. Benjamin and P.M. Hayden, they showed
that multiplayer quantum games can exhibit certain forms of pure quantum
equilibrium that have no analogue in classical games, or even in two player
quantum games \cite{6}. Although most of the works are focused on maximally
entangled quantum games, game of varying entanglement is also investigated
\cite{8,15}. For the particular case of the two-player quantum Prisoners'
Dilemma, two thresholds for the game's entanglement is found, and the
phenomena which are very much like phase transitions\ are also revealed. Even
though quantum game are played mostly on paper, the first experimental
realization of quantum games has also been implemented on a NMR quantum
computer \cite{10}.

In this paper, we investigate the phase-transition-like behavior
of quantum games, using a proposed method which would help to
illuminate the origin of such kind of behavior. For the
generalized version of Prisoners' Dilemma, we find that, with
different settings of the numerical values for the payoff table,
even though the classical game behaves the same, the quantum game
behaves greatly differently and exhibits interesting
phase-transition-like behavior in the entanglement-payoff diagram.
We find thresholds for the amount of entanglement that separate
different regions for the game. The properties of the game changes
discontinuously when its entanglement goes across these
thresholds, creating the phase-transition-like behavior. We
present investigation for both the case where the strategic space
is restricted as in Ref. \cite{4} and the case where the players
are allowed to adopt any unitary operations as their strategies.
In the case where the strategic space is restricted, the
phase-transition-like behavior exhibits interesting variation with
respect to the change of the numerical values in the payoff table,
so does the property of the game. In the case where the players
are allowed to adopt any unitary operations, the game has an
boundary, being a function of the numerical values in the payoff
table, for its entanglement. The quantum game has an infinite
number of Nash equilibria if its entanglement is below the
boundary, otherwise no pure strategic Nash equilibrium could be
found when its entanglement exceeds the boundary.

The proposed method would help to illuminate the origin of such
kind of phase-transition-like behavior. In this method, strategies
of players are corresponding to unit vectors in some real space,
and the searching for Nash equilibria includes a procedure of
finding the eigenvector of some matrix that corresponds to the
maximal eigenvalue. In the particular case presented in this
paper, the eigenvalues are functions of the amount of
entanglement, and thus there can be an eigenvalue-crossing.
Crossing an eigenvalue-crossing point makes the eigenvector that
corresponds to the maximal eigenvalue changes discontinuously,
indicating the discontinuous change of the properties of the
quantum game, as well as the phase-transition-like behavior.

\section{\label{sec2}Quantization of The Generalized Prisoners' Dilemma}

\begin{table}[b]
\caption{The general form of the Prisoners' Dilemma. The first
entry in the parenthesis denotes the payoff of Alice and the
second number the payoff of Bob. The entries in this table should
satisfy conditions: $t>r>p>s$ (see in Reference \protect\cite{9}).
The meanings of the symbols in the table is as follows. $C$:
Cooperate; $D$: Defect; $r$: reward; $p$: punishment; $t$:
temptation; $s$: sucker's payoff.}
\label{Table1}%
\begin{indented}
\item[]\begin{tabular}{ccc}
\br & Bob: $C$ & Bob: $D$ \\
\mr
Alice: $C$ & $\left( r ,r \right) $ & $\left( s ,t
\right) $ \\
Alice: $D$ & $\left( t ,s \right) $ & $\left( p ,p \right) $ \\
\br
\end{tabular}
\end{indented}
\end{table}

The classical Prisoners' Dilemma is the most widely studied and used paradigm
as a non-zero-sum game that could have an equilibrium outcome which is unique,
but fails to be Pareto optimal. The importance of this game lies in the fact
that many social phenomena with which we are familiar seem to have Prisoner's
Dilemma at their core. The general form of the Prisoners' Dilemma \cite{9} is
shown as in Table \ref{Table1}, with suggestive names for the strategies and
payoffs. The condition $t>r>p>s$ guarantees that strategy $D$ dominates
strategy $C$ for both players, and that the unique equilibrium at $\left(
D,D\right)  $ is Pareto inferior to $\left(  C,C\right)  $.%

\begin{figure}[tbp]
\begin{indented}
\item[]\epsfbox{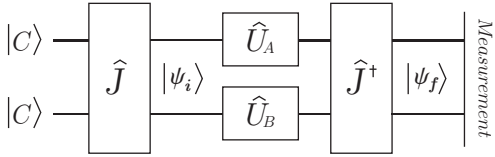}
\end{indented}
\caption{\label{fig1}The physical model for two player quantum Prisoners' Dilemma.}
\end{figure}

The physical model of the quantum Prisoners' Dilemma is originally proposed by
J. Eisert \textit{et al.} as shown in Fig. \ref{fig1}. Together with the
payoff table for the general Prisoners' Dilemma, the scheme can represent the
generalized quantum\ Prisoners' Dilemma. In this scheme the game has two
qubits, one for each player. The possible outcomes of the classical strategies
$D$ and $C$ are assigned to two basis $\left\vert D\right\rangle $ and
$\left\vert C\right\rangle $ in the Hilbert space of a qubit. Hence the state
of the game at each instance is described by a vector in the tensor product
space which is spanned by the classical game basis $\left\vert CC\right\rangle
$, $\left\vert CD\right\rangle $, $\left\vert DC\right\rangle $ and
$\left\vert DD\right\rangle $, where the first and second entries refer to
Alice's and Bob's qubits respectively. The initial state of the game is given
by%
\begin{equation}
\left\vert \psi_{i}\right\rangle =\hat{J}\left\vert CC\right\rangle
,\label{eq 1}%
\end{equation}
where $\hat{J}$\ is a unitary operator which is known to both players.
Strategic moves of Alice and Bob are associated with unitary operators
$\hat{U}_{A}$ and $\hat{U}_{B}$ respectively, which are chosen from a
strategic space $S$. At the final stage, the state of the game is%
\begin{equation}
\left\vert \psi_{f}\right\rangle =\hat{J}^{\dag}\left(  \hat{U}_{A}\otimes
\hat{U}_{B}\right)  \hat{J}\left\vert CC\right\rangle .\label{eq 2}%
\end{equation}
The subsequent measurement yields a particular result and the expect payoffs
of the players are given by%
\begin{equation}
\left\{
\begin{array}
[c]{c}%
\$_{A}=rP_{CC}+pP_{DD}+tP_{DC}+sP_{CD}\\
\$_{B}=rP_{CC}+pP_{DD}+sP_{DC}+tP_{CD}%
\end{array}
\right.  ,\label{eq 3}%
\end{equation}
where $P_{\sigma\tau}=\left\vert \left\langle \sigma\tau\right\vert \left.
\psi_{f}\right\rangle \right\vert ^{2}$ $\left(  \sigma,\tau\in\left\{
C,D\right\}  \right)  $ is the probability that $\left\vert \psi
_{f}\right\rangle $\ collapses into basis $\left\vert \sigma\tau\right\rangle
$.

In the general case, strategies for players could be any unitary operations.
However, since the overall phase factor of $\left\vert \psi_{f}\right\rangle $
will not affect the final results of the game, we can safely set the strategic
space $S=SU\left(  2\right)  $ as in Refs. \cite{4} and \cite{6}, without loss
of generality.

As we known, an operator $\hat{U}\in SU\left(  2\right)  $ can be written as%
\begin{equation}
\hat{U}=w\cdot\hat{I}_{2}+x\cdot i\hat{\sigma}_{x}+y\cdot i\hat{\sigma}%
_{y}+z\cdot i\hat{\sigma}_{z},\label{eq 5}%
\end{equation}
with $w,x,y,z\in\left[  -1,1\right]  $ and $w^{2}+x^{2}+y^{2}+z^{2}=1$. This
enables us to represent $\hat{U}$ directly by a four-dimensional real vector%
\begin{equation}
u=\left(  w,x,y,z\right)  \in\mathbb{R}^{4},
\end{equation}
with $u\cdot u^{T}=w^{2}+x^{2}+y^{2}+z^{2}=1$ (superscript $T$ denotes
\textit{Transpose}), and its components are denoted as $u^{1}=w,u^{2}%
=x,u^{3}=y,u^{4}=z$.

Denote Alice's strategy by $u_{A}$ and Bob's by $u_{B}$, the payoffs in Eq.
(\ref{eq 3}) can be written as%
\begin{equation}
\left\{
\begin{array}
[c]{c}%
\$_{A}=\$_{A}\left(  u_{A},u_{B}\right)  =\sum_{ij,kl}\$_{ij,kl}^{A}\cdot
u_{A}^{i}u_{A}^{j}u_{B}^{k}u_{B}^{l}\\
\$_{B}=\$_{B}\left(  u_{A},u_{B}\right)  =\sum_{ij,kl}\$_{ij,kl}^{B}\cdot
u_{B}^{i}u_{B}^{j}u_{A}^{k}u_{A}^{l}%
\end{array}
\right.  ,\label{eq 14}%
\end{equation}
where $i,j,k,l$ run from $1$ to $4$, $\left(  \$_{ij,kl}^{A}\right)  $\ and
$\left(  \$_{ij,kl}^{B}\right)  $\ are certain tensors. The formulation of
$\left(  \$_{ij,kl}^{A}\right)  $ and $\left(  \$_{ij,kl}^{B}\right)  $ in Eq.
(\ref{eq 14}) are not uniquely determined. However if restricted to be
symmetric, \textit{i.e.} $\$_{ij,kl}^{A}=\$_{ji,kl}^{A}=\$_{ij,lk}^{A}$ and
$\$_{ij,kl}^{B}=\$_{ji,kl}^{B}=\$_{ij,lk}^{B}$ (this can always be done), they
both can be uniquely determined. The calculations for $\left(  \$_{ij,kl}%
^{A}\right)  $\ and $\left(  \$_{ij,kl}^{B}\right)  $\ could be
found in \ref{app1}. Eqs. (\ref{eq 14}) are actually very general
formulations for any static quantum game expressed as in Table
\ref{Table1} and Fig. \ref{fig1} (the gate $\hat{J}^{\dag}$ prior
to measurement can even be replaced by other unitary
transformation, not necessarily the inverse of $\hat{J}$). All the
structural information of the game, including the classical payoff
table and the physical model, is represented by the tensors
$\left(  \$_{ij,kl}^{A}\right)  $ and $\left(
\$_{ij,kl}^{B}\right)  $. In the Prisoners' Dilemma, we have
$\$_{ij,kl}^{A}\equiv\$_{ij,kl}^{B}$ due to the symmetric
structure of the game. In an asymmetric game, $\left(
\$_{ij,kl}^{A}\right)  $ does not necessarily equals $\left(  \$_{ij,kl}%
^{B}\right)  $.

Defining $\$_{ij,kl}\equiv\$_{ij,kl}^{A}\equiv\$_{ij,kl}^{B}$, Eq.
(\ref{eq 14}) can be re-expressed as%

\begin{equation}
\left\{
\begin{array}
[c]{c}%
\$_{A}\left(  u_{A},u_{B}\right)  =\sum_{ij}\left(  \sum_{kl}\$_{ij,kl}\cdot
u_{B}^{k}u_{B}^{l}\right)  u_{A}^{i}u_{A}^{j}=u_{A}\cdot P\left(
u_{B}\right)  \cdot u_{A}^{T}\\
\$_{B}\left(  u_{A},u_{B}\right)  =\sum_{ij}\left(  \sum_{kl}\$_{ij,kl}\cdot
u_{A}^{k}u_{A}^{l}\right)  u_{B}^{i}u_{B}^{j}=u_{B}\cdot P\left(
u_{A}\right)  \cdot u_{B}^{T}%
\end{array}
\right.  ,\label{eq 15}%
\end{equation}
where $P\left(  u\right)  $ is a symmetric matrix as a function of $u$, whose
$i,j$-th element satisfies%
\begin{equation}
\left(  P\left(  u\right)  \right)  _{ij}=\sum_{kl}\$_{ij,kl}\cdot u^{k}%
u^{l}.\label{eq 16}%
\end{equation}

Let $\left(  u_{A}^{\ast},u_{B}^{\ast}\right)  $ be a Nash equilibrium of the
game, we can see that, from Eq. (\ref{eq 15}), $u_{A}\cdot P\left(
u_{B}^{\ast}\right)  \cdot u_{A}^{T}$ reaches its maximum at $u_{A}%
=u_{A}^{\ast}$ and simultaneously $u_{B}\cdot P\left(  u_{A}^{\ast}\right)
\cdot u_{B}^{T}$ reaches its maximum at $u_{B}=u_{B}^{\ast}$. In terms of game
theory, we say that $u_{A}^{\ast}$ dominates $u_{B}^{\ast}$ and $u_{B}^{\ast}$
dominates $u_{A}^{\ast}$. Together with $u_{A}^{\ast}\cdot\left(  u_{A}^{\ast
}\right)  ^{T}=u_{B}^{\ast}\cdot\left(  u_{B}^{\ast}\right)  ^{T}=1$, we can
conclude that $u_{A}^{\ast}$ ($u_{B}^{\ast}$) must be the eigenvector of
$P\left(  u_{B}^{\ast}\right)  $ [$P\left(  u_{A}^{\ast}\right)  $] which
corresponds to the maximal eigenvalue, and the corresponding eigenvalue is
exactly the payoff for Alice (Bob) at this Nash equilibrium. This analysis
also tells that the dominant strategy against a given strategy $u$ must be the
eigenvector of $P\left(  u\right)  $ that corresponds to the maximal eigenvalue.

In the following, we will first investigate the general Prisoners'
Dilemma in the case that the strategic space is restricted to be
the 2-parameter subset of $SU\left(  2\right)  $ as given in Ref.
\cite{4}. Then we investigate this game when the players are
allowed to adopt any unitary strategic operations. Here we shall
note that some authors \cite{17} have argued that the restriction
on the strategic space given in Ref. \cite{4} has no physical
basis, and it does restrict generality. However, apart from these
arguments, it is still an interesting case and a good instance to
show how the phase-transition-like behavior originates. Yet the
particular results achieved hold only for this very specific set
of strategies.

\section{\label{sec3}Two-Parameter Set of Strategies}

In the case of two-parameter set of strategies, the strategic space $S$ is
restricted to the two-parameter subset of $SU\left(  2\right)  $ as follows
\cite{4},%
\begin{equation}
\hat{U}\left(  \theta,\varphi\right)  =\left(
\begin{array}
[c]{cc}%
e^{i\varphi}\cos\theta/2 & \sin\theta/2\\
-\sin\theta/2 & e^{-i\varphi}\cos\theta/2
\end{array}
\right)  ,\label{eq 4}%
\end{equation}
with $\theta\in\left[  0,\pi\right]  $ and $\varphi\in\left[  0,\pi/2\right]
$.

As illustrated in details by J. Eisert \textit{et al. }\cite{4}, in order to
guarantee that the classical Prisoners' Dilemma is faithfully represented, the
form of $\hat{J}$ should be%
\begin{equation}
\hat{J}=e^{i\gamma\hat{D}\otimes\hat{D}/2}=\cos\frac{\gamma}{2}\hat{C}%
\otimes\hat{C}+i\sin\frac{\gamma}{2}\hat{D}\otimes\hat{D},\label{eq 7}%
\end{equation}
where $\hat{C}=$ $\hat{U}\left(  0,0\right)  $, $\hat{D}=\hat{U}\left(
\pi,0\right)  $,\ and $\gamma\in\left[  0,\pi/2\right]  $ is in fact a measure
for the game's entanglement.

Eq. (\ref{eq 4}) can be rewritten as%
\begin{eqnarray}
\hat{U}\left(  \theta,\varphi\right)   & =\cos\frac{\theta}{2}\cos\varphi
\cdot\hat{I}_{2}+\sin\frac{\theta}{2}\cdot i\hat{\sigma}_{y}+\cos\frac{\theta
}{2}\sin\varphi\cdot i\hat{\sigma}_{z}\nonumber\\
& =w\cdot\hat{I}_{2}+y\cdot i\hat{\sigma}_{y}+z\cdot i\hat{\sigma}%
_{z},\label{eq 12}%
\end{eqnarray}
where $w=\cos\frac{\theta}{2}\cos\varphi,y=\sin\frac{\theta}{2},z=\cos
\frac{\theta}{2}\sin\varphi$. Obviously we have $w,y,z\in\left[  0,1\right]  $
and $\hat{U}\left(  \theta,\varphi\right)  \in SU\left(  2\right)  $ implies
that $w^{2}+y^{2}+z^{2}=1$. Since $\hat{U}\left(  \theta,\varphi\right)  $ and
$-\hat{U}\left(  \theta,\varphi\right)  $ represent the same strategy, it is
enough to restrict ourselves with $w,y,z\in\left[  -1,1\right]  $. Therefore
in the case of two-parameter set of strategies, $\hat{U}\left(  \theta
,\varphi\right)  $ can be represented by a three-dimensional real vector%
\begin{equation}
u=\left(  w,y,z\right)  \in\mathbb{R}^{3},
\end{equation}
with $u\cdot u^{T}=w^{2}+y^{2}+z^{2}=1$. Eqs. (\ref{eq 14}, \ref{eq 15},
\ref{eq 16}) will remain their form, except that all the indices run only from
$1 $ to $3$, rather than from $1$ to $4$. Obviously we have $\hat{C}%
\sim\left(  1,0,0\right)  ,\hat{D}\sim\left(  0,1,0\right)  ,\hat{Q}%
\sim\left(  0,0,1\right)  $, in which \textquotedblleft$\sim$%
\textquotedblright\ means \textquotedblleft represent (by)\textquotedblright.
In the remaining part of this paper, we do not distinguish a unitary operator
and the corresponding vector (3-dimensional or 4-dimensional), as long as
there is not ambiguity.

In Ref. \cite{8}, we investigated this game in the case that $\left(
r,p,t,s\right)  =\left(  3,1,5,0\right)  $ and observed the phenomenon that
are very much like phase transitions. In the generalized quantum Prisoners'
Dilemma, such phase-transition-like behavior still exists. In fact, there
exist two thresholds for the game's entanglement, $\gamma_{th1}=\arcsin
\sqrt{\left(  p-s\right)  /\left(  t-s\right)  }$ and $\gamma_{th2}%
=\arcsin\sqrt{\left(  t-r\right)  /\left(  t-s\right)  }$. We hereby prove
that, for $0\leqslant\gamma<\gamma_{th1}$, the strategic profile $\hat
{D}\otimes\hat{D}$ is the Nash equilibrium with payoffs $\$_{A}=\$_{B}=p$. For
$\gamma_{th2}<\gamma\leqslant\pi/2$, the strategic profile $\hat{Q}\otimes
\hat{Q}$ is the Nash equilibrium with payoffs $\$_{A}=\$_{B}=r$. If
$\gamma_{th1}<\gamma_{th2}$ and $\gamma_{th1}\leqslant\gamma\leqslant
\gamma_{th2}$, the game has two Nash equilibria $\hat{D}\otimes\hat{Q}$ and
$\hat{Q}\otimes\hat{D}$. The payoff for the player who adopts $\hat{D}$ is
$s+\left(  t-s\right)  \cos^{2}\gamma$ while for the player who adopts
$\hat{Q}$\ is $s+\left(  t-s\right)  \sin{}^{2}\gamma$. While if $\gamma
_{th2}<\gamma_{th1}$ and $\gamma_{th2}\leqslant\gamma\leqslant\gamma_{th1}$,
both $\hat{D}\otimes\hat{D}$\ and $\hat{Q}\otimes\hat{Q}$\ are Nash equilibria
of the game. We obtain these conclusions through the following steps:

Assume one player adopts strategy $\hat{D}$, the payoff for the other as the
function of his/her strategy $u$ is%
\begin{equation}
u\cdot P\left(  \hat{D}\right)  \cdot u^{T},
\end{equation}
where the explicit expression of $P\left(  \hat{D}\right)  $\ is (the
calculation could be found in \ref{app2})%
\begin{equation}
P\left(  \hat{D}\right)  =\left(
\begin{array}
[c]{ccc}%
s & 0 & 0\\
0 & p & 0\\
0 & 0 & s+\left(  t-s\right)  \sin{}^{2}\gamma
\end{array}
\right)  .\label{eq 8}%
\end{equation}
If $0\leqslant\gamma<\gamma_{th1}=\arcsin\sqrt{\left(  p-s\right)  /\left(
t-s\right)  }$, the maximal eigenvalue of $P\left(  \hat{D}\right)  $ is $p$,
and the corresponding eigenvector is $\left(  0,1,0\right)  \sim\hat{D}$. If
$\gamma_{th1}<\gamma\leqslant\pi/2$, the maximal eigenvalue of $P\left(
\hat{D}\right)  $ is $s+\left(  t-s\right)  \sin{}^{2}\gamma$, and the
corresponding eigenvector is $\left(  0,0,1\right)  \sim\hat{Q}$. Therefore
$\hat{D}$ dominates $\hat{D}$ for $0\leqslant\gamma<\gamma_{th1}$ while
$\hat{Q}$ dominates $\hat{D}$ for $\gamma_{th1}<\gamma\leqslant\pi/2$. For the
same time we have $\$_{A}\left(  \hat{D},\hat{D}\right)  =\$_{B}\left(
\hat{D},\hat{D}\right)  =p$ and $\$_{A}\left(  \hat{Q},\hat{D}\right)
=\$_{B}\left(  \hat{D},\hat{Q}\right)  =s+\left(  t-s\right)  \sin{}^{2}%
\gamma$.

While assume one player adopts strategy $\hat{Q}$, the payoff for the other as
the function of his/her strategy $u$ is%
\begin{equation}
u\cdot P\left(  \hat{Q}\right)  \cdot u^{T},
\end{equation}
where the explicit expression of $P\left(  \hat{Q}\right)  $\ is (the
calculation could be found in \ref{app2})%
\begin{equation}
P\left(  \hat{Q}\right)  =\left(
\begin{array}
[c]{ccc}%
r-\left(  r-p\right)  \sin{}^{2}\gamma & 0 & 0\\
0 & t-\left(  t-s\right)  \sin{}^{2}\gamma & 0\\
0 & 0 & r
\end{array}
\right)  .\label{eq 9}%
\end{equation}
If $0\leqslant\gamma<\gamma_{th2}=\arcsin\sqrt{\left(  t-r\right)  /\left(
t-s\right)  }$, the maximal eigenvalue of $P\left(  \hat{Q}\right)  $ is
$t-\left(  t-s\right)  \sin{}^{2}\gamma$, and the corresponding eigenvector is
$\left(  0,1,0\right)  \sim\hat{D}$. If $\gamma_{th2}<\gamma\leqslant\pi/2$,
the maximal eigenvalue of $P\left(  \hat{Q}\right)  $ is $r$, and the
corresponding eigenvector is $\left(  0,0,1\right)  \sim\hat{Q}$. Therefore
$\hat{D}$ dominates $\hat{Q}$ for $0\leqslant\gamma<\gamma_{th2}$ while
$\hat{Q}$ dominates $\hat{Q}$ for $\gamma_{th2}<\gamma\leqslant\pi/2$. For the
same time we have $\$_{A}\left(  \hat{Q},\hat{Q}\right)  =\$_{B}\left(
\hat{Q},\hat{Q}\right)  =r$ and $\$_{A}\left(  \hat{D},\hat{Q}\right)
=\$_{B}\left(  \hat{Q},\hat{D}\right)  =t-\left(  t-s\right)  \sin{}^{2}%
\gamma=s+\left(  t-s\right)  \cos^{2}\gamma$.

From the above analysis, we can see that when $0\leqslant\gamma<\gamma_{th1}$,
$\hat{D}\otimes\hat{D}$ is a Nash equilibrium of the game, and when
$\gamma_{th2}<\gamma\leqslant\pi/2$, $\hat{Q}\otimes\hat{Q}$ is a Nash
equilibrium of the game. If $\gamma_{th1}<\gamma_{th2}$ and $\gamma
_{th1}\leqslant\gamma\leqslant\gamma_{th2}$, $\hat{D}$ dominates $\hat{Q} $
and $\hat{Q}$ dominates $\hat{D}$, hence both $\hat{D}\otimes\hat{Q}$ and
\ $\hat{Q}\otimes\hat{D}$ are Nash equilibria of the game. While if
$\gamma_{th2}<\gamma_{th1}$ and $\gamma_{th2}\leqslant\gamma\leqslant
\gamma_{th1}$, both $\hat{D}\otimes\hat{D}$\ and $\hat{Q}\otimes\hat{Q}$\ are
Nash equilibria of the game. The corresponding payoffs are also obtained.%

\begin{figure}[tbp]
\begin{indented}
\item[]\epsfbox{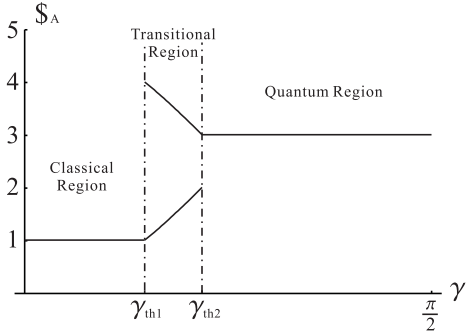}
\end{indented}
\caption{\label{fig2}The payoff function of Alice with respect to the amount of the
entanglement in the case of two-parameter strategies. The numerical values in
the payoff matrix are set as $(r=3,p=1,t=5,s=0)$ such that $r+p<t+s$. The
region between two thresholds are the transitional region from classical to
quantum, in which the game has two asymmetric Nash equilibria although the
game is symmetric with respect to the interchange of the players.}
\end{figure}

In the case that the entries in the payoff table are taken as $\left(
r=3,p=1,t=5,s=0\right)  $, which has been investigated in Ref\cite{8}, the
game has two thresholds for the amount of the game's entanglement. Due to the
two thresholds, the game is divided into three regions, the classical region,
the quantum region, and the transitional region from classical to quantum. In
the general quantum Prisoners' Dilemma, there still exist two thresholds and
the phase-transition-like behavior shows up again. However the situation may
be more complicated because the two thresholds have no deterministic relations
in magnitude. In fact, the case that $\left(  r=3,p=1,t=5,s=0\right)  $ is
just an instance of the more general case of $r+p<t+s$. For the game under
this condition, it is obviously that $\gamma_{th1}<\gamma_{th2}$ and the game
behaves similarly to the one with $\left(  r=3,p=1,t=5,s=0\right)  $. Fig.
\ref{fig2} depicts the payoff of Alice as the function of $\gamma$ when both
players resort to Nash equilibrium in the case of $r+p<t+s$. In the
transitional regions, the two Nash equilibria are fully equivalent. Since
there is no communication between two players, one player will have no idea
which equilibrium strategy the other player chooses. So the strategy mismatch
situation will probably occur. A more severe problem is that, since strategy
$\hat{D}$ will lead to a better payoff so both players will be tempted to
choose $\hat{D}$ and the final payoff for both of them will become $p$, which
happens to be the catch of the dilemma in the classical game.

\begin{figure}[tbp]
\begin{indented}
\item[]\epsfbox{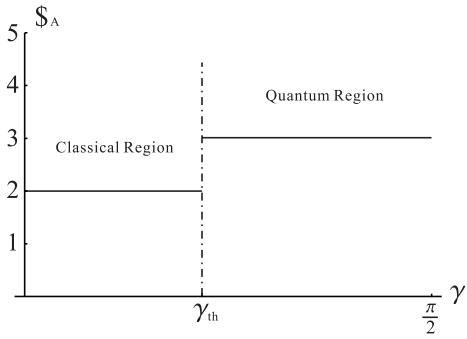}
\end{indented}
\caption{\label{fig3}The payoff function of Alice with respect to the amount of the
entanglement in the case of two-parameter strategies. The numerical values in
the payoff matrix are set as $(r=3,p=2,t=5,s=0)$ such that $r+p=t+s$. The two
thresholds converge to be a unique one $\gamma_{th}$ and the transitional
region no longer exists.}
\end{figure}

An interesting situation is, as we can see, if $\gamma_{th1}=\gamma_{th2}$,
the transitional region will disappear. The condition $\gamma_{th1}%
=\gamma_{th2}$ implies that%
\begin{equation}
r+p=t+s.\label{eq 10}%
\end{equation}
Note that we should keep in mind that the basic condition $t>r>p>s$ must be
satisfied to maintain the properties of the classical game.\ And under the
condition in Eq. (\ref{eq 10}) the game has only one threshold for its
entanglement $\gamma_{th}=\gamma_{th1}=\gamma_{th2}$. Hence the game exhibits
only two regions, one is classical and the other is quantum. The transitional
region in which the game has two asymmetric Nash equilibrium disappears. Under
the conditions $r+p=t+s$ and $t>r>p>s$, we plot the payoff of Alice as the
function of $\gamma$ in Fig. \ref{fig3} when both players resort to Nash equilibrium.%

\begin{figure}[tbp]
\begin{indented}
\item[]\epsfbox{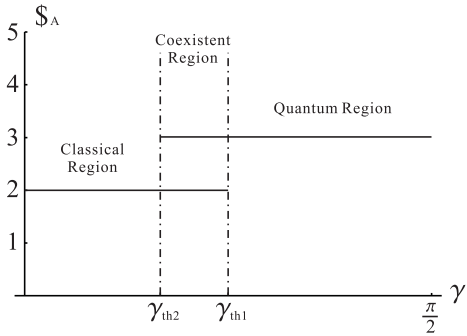}
\end{indented}
\caption{\label{fig4}The payoff function of Alice with respect to the amount of the
entanglement in the case of two-parameter strategies. The numerical values in
the payoff matrix are set as $(r=3,p=2,t=4,s=0)$ such that $r+p>t+s$. The
coexistent region emerges, in which both $\hat{D}\otimes\hat{D}$ and $\hat
{Q}\otimes\hat{Q}$ are Nash equilibria.}
\end{figure}

Now we consider what would happen in the game of $r+p>t+s$. In this case, we
have $\gamma_{th1}>\gamma_{th2}$. Therefore the game has no transitional
region, hence none of $\hat{D}\otimes\hat{Q}$ and $\hat{Q}\otimes\hat{D}$ is a
Nash equilibrium of the game. However both $\hat{D}\otimes\hat{D}$ and
$\hat{Q}\otimes\hat{Q}$ are still Nash equilibria in the region $\gamma
_{th2}\leqslant\gamma\leqslant\gamma_{th1}$. So for $\gamma_{th2}%
\leqslant\gamma\leqslant\gamma_{th1}$, a new region --- coexistent
region --- arises with two Nash equilibria. These two Nash
equilibria are both symmetric with respect to the interchange of
the two players. In this case, we illustrate the payoff of Alice
as the function of $\gamma$ in Fig. \ref{fig4}. We should also
note that in this case the multiple Nash equilibria brings a
situation different to that in the transitional regions with
$r+p<t+s$. The two Nash equilibria are not equivalent and
$\hat{Q}\otimes\hat{Q}$ gives higher payoffs to both players than
does $\hat{D}\otimes\hat{D}$. Therefore it is a quite reasonable
assumption that the players are most likely to resort to the
equilibrium $\hat{Q}\otimes\hat{Q}$ rather than
$\hat{D}\otimes\hat{D}$, since they are both trying to maximize
their individual payoffs. However, one still can not claim that
the players will definitely resort to the equilibrium that gives
higher payoffs. But if they do, the final results of the game will
then be the same as in the quantum region with
$\gamma>\gamma_{th1}$, and the dilemma will be resolved.

An interesting question is that can the game behave full quantum-mechanically
no matter how much it is entangled for some particular numerical value of
$\left(  r,p,t,s\right)  $, \textit{i.e.} have only the quantum region
(without the presence of classical, transitional or coexistent regions). If it
can, we immediately deduce that $\gamma_{th2}=0$. This means $t=r$, which
contradicts the basic condition $t>r>p>s$. Hence the game cannot always have
$\hat{Q}\otimes\hat{Q}$ as its equilibrium in the whole domain of $\gamma$
from $0$ to $\pi/2$, as long as the game remains a \textquotedblleft
Prisoners' Dilemma\textquotedblright. In fact, as long as the condition
$t>r>p>s$ holds, none of $\gamma_{th1}$ and $\gamma_{th2}$ could reach $0$ or
$\pi/2$, hence none of the classical and quantum regions will disappear.%

\section{\label{sec4}General Unitary Operations}

In this section, we investigate the generalized quantum Prisoners' Dilemma
when both the players can access to any unitary operations as their
strategies, rather than in a restricted subset in Eq. (\ref{eq 4}). The method
for analyzing is clearly described in section \ref{sec2}. The result is that,
there exist a boundary $\gamma_{B}=\arcsin\sqrt{\left(  p-s\right)  /\left(
p+t-r-s\right)  }$ for the game's entanglement. If $\gamma<\gamma_{B}$, there
are infinite Nash equilibrium. Any strategic profile $\left\{  \left(
0,\alpha,\beta,0\right)  ,\left(  0,\beta,\alpha,0\right)  \right\}  $
($\alpha^{2}+\beta^{2}=1$) is a Nash equilibrium. Each of them results in the
same payoffs $\$_{A}=\$_{B}=p+\left(  r-p\right)  \sin^{2}\gamma$. While as
long as $\gamma>\gamma_{B}$, there will be no Nash equilibrium for the game.
We prove these results as follows.

For the strategy $u_{1} = \left(  0,\alpha,\beta,0\right)  $ ($\alpha^{2}+\beta^{2}%
=1$), we have (the calculation could be found in \ref{app1}), with
$\epsilon\equiv\sin^{2}\gamma$,%
\begin{equation}
\fl P\left( u_{1} \right) =\left(
\begin{array}
[c]{cccc}%
s+\left(  t-s\right)  \alpha^{2}\epsilon & 0 & 0 & \left(  s-t\right)
\alpha\beta\epsilon\\
0 & p+\left(  r-p\right)  \beta^{2}\epsilon & \left(  r-p\right)  \alpha
\beta\epsilon & 0\\
0 & \left(  r-p\right)  \alpha\beta\epsilon & p+\left(  r-p\right)  \alpha
^{2}\epsilon & 0\\
\left(  s-t\right)  \alpha\beta\epsilon & 0 & 0 & s+\left(  t-s\right)
\beta^{2}\epsilon
\end{array}
\right)  .\label{eq 17}%
\end{equation}
The eigenvalues and corresponding eigenvectors of $P\left(  \left(
0,\alpha,\beta,0\right)  \right)  $ in Eq. (\ref{eq 17}) are%

\begin{equation}
\left\{
\begin{array}
[c]{ll}%
p & \quad\left(  0,\alpha,-\beta,0\right) \\
s & \quad\left(  \beta,0,0,\alpha\right) \\
p+\left(  r-p\right)  \sin^{2}\gamma & \quad\left(  0,\beta,\alpha,0\right) \\
s+\left(  t-s\right)  \sin^{2}\gamma & \quad\left(  \alpha,0,0,-\beta\right)
\end{array}
\right.  .\label{eq 20}%
\end{equation}
If $\gamma<\gamma_{B}$, the maximal eigenvalue is $p+\left(  r-p\right)
\sin^{2}\gamma$ and the corresponding eigenvector is $\left(  0,\beta
,\alpha,0\right)  $. Therefore $\left(  0,\beta,\alpha,0\right)  $ dominates
$\left(  0,\alpha,\beta,0\right)  $, and vice versa (by exchanging $\alpha$
and $\beta$ in Eqs. (\ref{eq 17}, \ref{eq 20})). Hence any strategic profile
$\left\{  \left(  0,\alpha,\beta,0\right)  ,\left(  0,\beta,\alpha,0\right)
\right\}  $ ($\alpha^{2}+\beta^{2}=1$) is a Nash equilibrium.

While if $\gamma>\gamma_{B}$, the dominant strategy against
$\left( 0,\alpha,\beta,0\right)  $ turns to be $\left(
\alpha,0,0,-\beta\right)  $. For the strategy $ u_{2} = \left(
\alpha,0,0,-\beta\right)  $, we have (the calculation could be
found in \ref{app1}), with $\epsilon\equiv
\sin^{2}\gamma$,%
\begin{equation}
\fl P\left(  u_{2}  \right)  =\left(
\begin{array}
[c]{cccc}%
r+\left(  p-r\right)  \beta^{2}\epsilon & 0 & 0 & \left(  r-p\right)
\alpha\beta\epsilon\\
0 & t+\left(  s-t\right)  \alpha^{2}\epsilon & \left(  s-t\right)  \alpha
\beta\epsilon & 0\\
0 & \left(  s-t\right)  \alpha\beta\epsilon & t+\left(  s-t\right)  \beta
^{2}\epsilon & 0\\
\left(  r-p\right)  \alpha\beta\epsilon & 0 & 0 & r+\left(  p-r\right)
\alpha^{2}\epsilon
\end{array}
\right)  .\label{eq 18}%
\end{equation}
And the eigenvalues and corresponding eigenvectors of $P\left(  \left(
\alpha,0,0,-\beta\right)  \right)  $ in Eq. (\ref{eq 18}) are%

\begin{equation}
\left\{
\begin{array}
[c]{ll}%
r & \quad\left(  \alpha,0,0,\beta\right) \\
t & \quad\left(  0,\beta,-\alpha,0\right) \\
r+\left(  p-r\right)  \sin^{2}\gamma & \quad\left(  \beta,0,0,-\alpha\right)
\\
t+\left(  s-t\right)  \sin^{2}\gamma & \quad\left(  0,\alpha,\beta,0\right)
\end{array}
\right.  .\label{eq 21}%
\end{equation}
In Eq. (\ref{eq 21}), $\left(  0,\beta,-\alpha,0\right)  $ always corresponds
to the maximal eigenvalue $t$. Therefore no matter what the amount of
entanglement is, $\left(  0,\beta,-\alpha,0\right)  $ always dominates
$\left(  \alpha,0,0,-\beta\right)  $. With further analysis combining Eq.
(\ref{eq 20}) and Eq. (\ref{eq 21}), we find that when $\gamma>\gamma_{B}$,
$\left(  \alpha,0,0,-\beta\right)  $ dominates $\left(  0,\alpha
,\beta,0\right)  $, $\left(  0,\beta,-\alpha,0\right)  $ dominates $\left(
\alpha,0,0,-\beta\right)  $, $\left(  \beta,0,0,\alpha\right)  $ dominates
$\left(  0,\beta,-\alpha,0\right)  $, and finally $\left(  0,\alpha
,\beta,0\right)  $ dominates $\left(  \beta,0,0,\alpha\right)  $. No pair of
them can form a Nash equilibrium. In fact, it can be proved that no pair of
strategies in the region of $\gamma>\gamma_{B}$ can form a pure Nash
equilibrium of the game. However the game remains to have mixed Nash
equilibria \cite{16}.

\begin{figure}[tbp]
\begin{indented}
\item[]\epsfbox{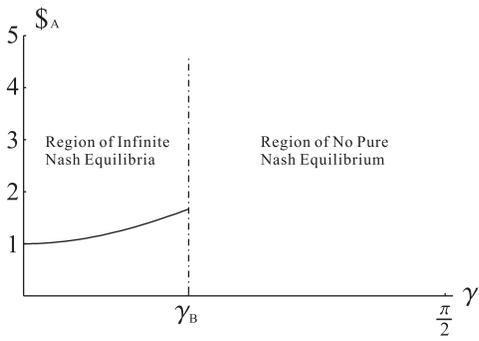}
\end{indented}
\caption{\label{fig5}The payoff function of Alice with respect to the amount of the
entanglement in the case that both players are allowed to adopt any unitary
operator as his/her strategy.}
\end{figure}

We depict the payoff function of Alice as a function of the amount of
entanglement when both players resort to Nash equilibrium (if there is one) in
Fig. \ref{fig5}. This figure also exhibits the phase-transition-like behavior
of the game. The boundary of entanglement divides the game into two regions:
in one of which the game has infinite Nash equilibria, while in the other the
game has no pure strategic Nash equilibrium.%

\section{Discussion and Conclusion}

In this paper, we investigate the discontinuous dependence of Nash
equilibria and payoffs on the game's entanglement for the general
quantum Prisoners' Dilemma. This discontinuity can be viewed as
the phase-transition-like behavior in the payoff-entanglement
diagram. We firstly investigate the generalized quantum Prisoners'
Dilemma when the strategic space is restricted to be a
two-parameter subset of $SU\left(  2\right)  $ as in Ref.
\cite{4}. With condition $r+p<t+s$, the game exhibits the
classical, quantum and transitional regions in its
payoff-entanglement diagram. The original Prisoners' Dilemma with
$\left(  r=3,p=1,t=5,s=0\right)  $ is just an instance for the
general game with condition $r+p<t+s$. In the classical region
$\hat{D}\otimes\hat{D}$ is the unique Nash equilibrium, and in the
quantum region the unique Nash equilibrium is
$\hat{Q}\otimes\hat{Q}$. While in the transitional region, two
asymmetric Nash equilibria, $\hat{D}\otimes\hat{Q}$ and
$\hat{Q}\otimes\hat{D}$, emerge, each leads to the asymmetric
result of the game in despite of the symmetry of the game itself.
If the entries in the payoff table satisfy that $r+p=t+s$, the
transitional region will disappear. The game has only one
threshold for the amount of its entanglement at which the game
transits from classical to quantum discontinuously. In the case
that $r+p>t+s$, a new region --- the coexistent region ---
emerges, replacing the transitional region. This new region is in
fact there where the classical region and the quantum region
overlap. In the coexistent region, the game has both
$\hat{D}\otimes\hat{D}$ and $\hat{Q}\otimes\hat{Q}$ as its Nash
equilibria. Since $\hat{Q}\otimes\hat{Q}$ is superior to
$\hat{D}\otimes \hat{D}$, one may expect both players most likely
to choose $\hat{Q}$\ as his/her strategy, and the dilemma will be
resolved if they do so. We also explored the phase-transition-like
behavior of the quantum game in the case where both players are
allowed to adopt any unitary transformations as their strategies.
The game has an boundary for its entanglement, being a function of
the numerical values in the payoff table, below which the game has
infinite Nash equilibria, while above which the game has no pure
strategic Nash equilibrium.

The phase-transition-like behavior presented in this paper is very
much like phase transitions in real physical systems \cite{14},
not only phenomenally but also mathematically. For a certain
physical system whose Hamiltonian is dependent of some parameter,
a special case is that the eigenfunctions of the Hamiltonian is
independent of the parameter even though the eigenvalues vary with
it. Then there can be a level-crossing where an excited level
becomes the ground state, creating a point of a non-analyticity of
the ground state energy as a function of the parameter, as well as
a discontinuous dependence of the ground state on the parameter. A
quantum phase transition is hence viewed as any point of
non-analyticity in the ground state energy of the system
concerned. In the generalized quantum Prisoners' Dilemma, the
dominant strategy against a given strategy $u$ is the eigenvector
that corresponds to the maximal eigenvalue of matrix $P\left(
u\right) $ (see in Section \ref{sec2}). Since $P\left(  u\right)
$\ is a function of the amount of entanglement $\gamma$, the
eigenvalues may cross. This eigenvalue-crossing makes the
eigenvector that corresponds to the maximal eigenvalue changes
discontinuously. It also creates a non-analyticity of the payoff
(the maximal eigenvalue) as a function of $\gamma$, and the game
exhibit phase-transition-like behavior. The method proposed in
this paper would help to illuminate the origin of the
phase-transition-like behavior of quantum games, and we hope it
would further help investigate quantum games more intensively, and
more profound results may be derived.

\ack
This work was supported by the Nature Science Foundation of
China (Grants No. 10075041 and No. 10075044), the National
Fundamental Research Program (Grant No. 2001CB309300) and the
ASTAR Grant No. 012-104-0040.

\appendix

\section{\label{app1}Calculations For General Unitary Operations}

Denote Alice's strategy by $u_{A}=\left(  u_{A}^{1},u_{A}^{2},u_{A}^{3}%
,u_{A}^{4}\right)  $ and Bob's by $u_{B}=\left(  u_{B}^{1},u_{B}^{2},u_{B}%
^{3},u_{B}^{4}\right)  $, then substitute Eq. (\ref{eq 5}) into Eq.
(\ref{eq 2}), we have%
\begin{eqnarray}
\fl \left\vert \psi_{f}\right\rangle   =&[\left(  u_{A}^{1}u_{B}^{1}-u_{A}%
^{4}u_{B}^{4}\right)  +i\left(  u_{A}^{4}u_{B}^{1}+u_{A}^{1}u_{B}^{4}\right)
\cos\gamma-\left(  u_{A}^{3}u_{B}^{2}+u_{A}^{2}u_{B}^{3}\right)  \sin
\gamma]\left\vert CC\right\rangle +\nonumber\\
\fl & [-\left(  u_{A}^{1}u_{B}^{3}+u_{A}^{4}u_{B}^{2}\right)  +i\left(  u_{A}%
^{1}u_{B}^{2}-u_{A}^{4}u_{B}^{3}\right)  \cos\gamma+\left(  u_{A}^{3}u_{B}%
^{4}-u_{A}^{2}u_{B}^{1}\right)  \sin\gamma]\left\vert CD\right\rangle
+\nonumber\\
\fl & [-\left(  u_{A}^{3}u_{B}^{1}+u_{A}^{2}u_{B}^{4}\right)  +i\left(  u_{A}%
^{2}u_{B}^{1}-u_{A}^{3}u_{B}^{4}\right)  \cos\gamma+\left(  u_{A}^{4}u_{B}%
^{3}-u_{A}^{1}u_{B}^{2}\right)  \sin\gamma]\left\vert DC\right\rangle
+\nonumber\\
\fl & [\left(  u_{A}^{3}u_{B}^{3}-u_{A}^{2}u_{B}^{2}\right)  -i\left(  u_{A}%
^{3}u_{B}^{2}+u_{A}^{2}u_{B}^{3}\right)  \cos\gamma+\left(  u_{A}^{4}u_{B}%
^{1}+u_{A}^{1}u_{B}^{4}\right)  \sin\gamma]\left\vert
DD\right\rangle .
\fl \label{eq app1}%
\end{eqnarray}
Since the game is symmetric with respect to the interchange of the players, we
have%
\begin{equation}
\$_{A}\left(  u_{A},u_{B}\right)  \equiv\$_{B}\left(  u_{B},u_{A}\right)
,\forall u_{A},u_{B}\in SU\left(  2\right)  .\label{eq app2}%
\end{equation}
and we can immediately see from Eq. (\ref{eq 14}) that%
\begin{equation}
\$_{ij,kl}^{A}\equiv\$_{ij,kl}^{B}, i,j=1,2,3,4.\label{eq app3}
\end{equation}
And $P\left(  u\right)  $ (in Eq. (\ref{eq 16})) is symmetric too. Therefore
we can define $\$_{ij,kl}\equiv\$_{ij,kl}^{A}\equiv\$_{ij,kl}^{B}$ for
convenience. Substitute Eq. (\ref{eq app1}) into Eqs. (\ref{eq 3},
\ref{eq 14}), we can find the non-zero elements of $\left(  \$_{ij,kl}\right)
$ are (with $\$_{ij,kl}=\$_{ji,kl}=\$_{ij,lk}=\$_{ji,lk}$)%
\begin{eqnarray}
\$_{11,11}  & =\$_{44,44}=r,\$_{11,33}=\$_{44,22}=s,\nonumber\\
\$_{22,22}  & =\$_{33,33}=p,\$_{22,44}=\$_{33,11}=t,\nonumber\\
\$_{11,22}  & =\$_{44,33}=s+\left(  t-s\right)  \sin^{2}\gamma,\$_{11,44}%
=\$_{44,11}=r+\left(  p-r\right)  \sin^{2}\gamma,\nonumber\\
\$_{22,11}  & =\$_{33,44}=t+\left(  s-t\right)  \sin^{2}\gamma,\$_{22,33}%
=\$_{33,22}=p+\left(  r-p\right)  \sin^{2}\gamma,\nonumber\\
\$_{12,13}  & =-\$_{34,24}=\frac{1}{2}\left(  s-r\right)
\sin\gamma
,\$_{12,24}=-\$_{34,13}=\frac{1}{2}\left(  t-p\right)  \sin\gamma,\nonumber\\
\$_{13,12}  & =-\$_{24,34}=\frac{1}{2}\left(  t-r\right)
\sin\gamma
,\$_{13,34}=-\$_{24,12}=\frac{1}{2}\left(  p-s\right)  \sin\gamma,\nonumber\\
\$_{14,14}  & =-\$_{23,23}=\frac{1}{2}\left(  p-r\right)  \sin^{2}%
\gamma,\$_{14,23}=-\$_{23,14}=\frac{1}{2}\left(  s-t\right)  \sin^{2}%
\gamma.\label{eq app4}%
\end{eqnarray}

For strategy $\left(  0,\alpha,\beta,0\right)  $, we see that%
\begin{eqnarray}
\left(  P\left(  \left(  0,\alpha,\beta,0\right)  \right)  \right)  _{ij}  &
=\left(  P\left(  \left(  0,\alpha,\beta,0\right)  \right)  \right)
_{ji}\nonumber\\
& =\alpha^{2}\$_{ij,22}+\beta^{2}\$_{ij,33}+\alpha\beta\left(  \$_{ij,23}%
+\$_{ij,32}\right) \nonumber\\
& =\alpha^{2}\$_{ij,22}+\beta^{2}\$_{ij,33}+2\alpha\beta\$_{ij,23}%
\label{eq app5}%
\end{eqnarray}
and for $\left(  \alpha,0,0,-\beta\right)  $ we have%
\begin{eqnarray}
\left(  P\left(  \left(  \alpha,0,0,-\beta\right)  \right)  \right)  _{ij}  &
=\left(  P\left(  \left(  \alpha,0,0,-\beta\right)  \right)  \right)
_{ji}\nonumber\\
& =\alpha^{2}\$_{ij,11}+\beta^{2}\$_{ij,44}-\alpha\beta\left(  \$_{ij,14}%
+\$_{ij,41}\right) \nonumber\\
& =\alpha^{2}\$_{ij,11}+\beta^{2}\$_{ij,44}-2\alpha\beta\$_{ij,14}%
.\label{eq app6}%
\end{eqnarray}
Therefore Eq. (\ref{eq 17}) and (\ref{eq 18}) are obtained.

\section{\label{app2}Calculations For Two-Parameter Strategic Space}

The two-parameter strategic space can be obtained by restricting
$u^{2}=x\equiv0$ in the general case ($u^{2}=x$ is the second
component of $u $, not its squared length). Therefore the
expressions for $\$_{ij,kl}$ can be obtained from Eqs. (\ref{eq
app4}), by excluding all elements containing the index $2$, and
then replacing index $3$ by $2$ and $4$ by $3$. Therefore for the
case of two-parameter strategic space, we have all the non-zero
elements
(with $\$_{ij,kl}=\$_{ji,kl}=\$_{ij,lk}=\$_{ji,lk}$) as follows.%
\begin{eqnarray}
\$_{11,11}  & =\$_{33,33}=r,\$_{11,22}=s,\$_{22,22}=p,\$_{22,11}=t,\nonumber\\
\$_{33,22}  & =s+\left(  t-s\right)  \sin^{2}\gamma,\$_{11,33}=\$_{33,11}%
=r+\left(  p-r\right)  \sin^{2}\gamma,\nonumber\\
\$_{22,33}  & =t+\left(  s-t\right)  \sin^{2}\gamma,\$_{13,13}=\frac{1}%
{2}\left(  p-r\right)  \sin^{2}\gamma,\nonumber\\
\$_{23,12}  & =\frac{1}{2}\left(  p-t\right)  \sin\gamma,\$_{12,23}=\frac
{1}{2}\left(  p-s\right)  \sin\gamma.
\end{eqnarray}
Since $\hat{D}\sim\left(  0,1,0\right)  $ and $\hat{Q}\sim\left(
0,0,1\right)  $, it is obvious to see that%
\begin{eqnarray}
\left(  P\left(  \hat{D}\right)  \right)  _{ij}  & =\left(  P\left(  \hat
{D}\right)  \right)  _{ji}=\$_{ij,22},\\
\left(  P\left(  \hat{Q}\right)  \right)  _{ij}  & =\left(  P\left(  \hat
{Q}\right)  \right)  _{ji}=\$_{ij,33},
\end{eqnarray}
with $i,j=1,2,3$. The expressions in Eqs. (\ref{eq 8}) and (\ref{eq 9}) are
hence obtained.


\section*{References}

\begin{thebibliography}{99}
\bibitem {1}P. Ball, Nature Science Update, 18 Oct. 1999.

\bibitem {2}I. Peteron, Science News \textbf{156}, 334 (1999).

\bibitem {3}G. Collins, Sci. Am. Jan. 2000.

\bibitem {4}J. Eisert \textit{et al.}, Phys. Rev. Lett. \textbf{83}, 3077 (1999).

\bibitem {5}L. Marinatto and T. Weber. Phys. Lett. A \textbf{272}. 291 (2000).

\bibitem {6}S.C. Benjamin and P.M. Hayden, Phys. Rev. A \textbf{64}, 030301(R) (2001).

\bibitem {11}Jiangfeng Du \textit{et al.}, Phys. Lett. A \textbf{302}, 229 (2002).

\bibitem {7}D.A. Meyer, Phys. Rev. Lett. \textbf{82}, 1052 (1999).

\bibitem {8}Jiangfeng Du \textit{et al.}, Phys. Lett. A \textbf{289}, 9 (2001).

\bibitem {15}A.P. Flitney and D. Abbott, quant-ph/0209121.

\bibitem {10}Jiangfeng Du \textit{et al.}, Phys. Rev. Lett. \textbf{88},
137902 (2002).

\bibitem {9}P.D. Straffin, \textit{Game Theory and Strategy} (The Mathematical
Association of America, 1993). The original general Prisoners' Dilemma has an
additional condition, $r>\left(  s+t\right)  /2$, besides $t>r>p>s$. This
additional condition guarantees that even in an iterated game, the players
would be at least as well off always playing $\left(  C,C\right)  $ as
alternating between $\left(  C,D\right)  $ and $\left(  D,C\right)  $. So the
strategy profile $\left(  C,C\right)  $ is Pareto optimal in both static and
iterated game. In this paper we focus on the study of static games, so it is
unnecessary for us to consider this additional condition.

\bibitem {14}S. Sachdev, \textit{Quantum Phase Transitions} (Cambridge
University Press, 1999).

\bibitem {16}J. Eisert and M. Wilkins, J. Mod. Opt. \textbf{47}, 2543 (2000).

\bibitem {17}S.C. Benjamin and P.M. Hayden, Phys. Rev. Lett. \textbf{87},
069801 (2001).
\end{thebibliography}
\end{document}